# 台灣 COVID-19 本土首萬例中性別、年齡和縣市地區的關聯：對數線性模型分析

# Association Among Gender, Age, and Region in Taiwan's First Ten Thousand COVID-19 Cases: A Log-linear-model Analysis

洪泰晟 [1]、黃禮珊 [2*]

Tai-Cheng Hung [1], Li-Shan Huang [2]

[1] 國立陽明交通大學統計學研究所

Institute of Statistics, College of Science, National Yang Ming Chiao Tung University, Hsinchu, Taiwan, R.O.C.

[2] 國立清華大學統計學研究所

Institute of Statistics, College of Science, National Tsing Hua University, Hsinchu, Taiwan, R.O.C.

[*] 通訊作者：黃禮珊

地址：新竹市光復路二段 101 號第三綜合大樓 8F

E-mail：lhuang@stat.nthu.edu.tw

電話: 03-5715131 Ext. 33183

簡略題目(running title)：台灣 COVID-19 首萬例中性別年齡地區的關聯

# 台灣 COVID-19 本土首萬例中性別、年齡和縣市地區的關聯：

## 對數線性模型分析

### 摘要

**目標:** 本研究關注台灣 2020 年 1 月 22 日至 2021 年 6 月 11 日的 11290 本土 Covid-19 確診案例，了解其中年齡、性別與所居住縣市之間的關聯。**方法:** 使用台灣疾病管制署開放資料，整理成三維度（性別、年齡、縣市地區）列聯表，年齡組別為 0-29、30-59、60 歲以上，地區組別有兩種: (1) 北北基、桃竹苗、中彰投、雲嘉南、高屏、宜花東與離島共 7 組，(2) 北北基、桃竹苗分開及其它地區組別共 12 組。統計方法使用對數線性模型，模型選擇指標為 BIC。**結果:** BIC 最小的是包含三個成對交互項的模型，在交互項效應上，30-59 歲的女性比男性多($p<0.001$); 60 歲以上則是男性比女性多($p=0.028$)。苗栗縣男性比女性多($p<0.001$)。台北市 30-59 歲($p=0.002$)、60 歲以上($p<0.001$)皆比 0-29 歲的多；新北市有類似的年齡效應($p=0.062$、$p=0.018$); 苗栗縣 60 歲以上的比 0-29 少($p<0.001$)，桃園市交互項皆不顯著。年齡主效應 30-59、60 歲以上與 0-29(基準)的差異皆顯著($p=0.002$、$p=0.046$)。**結論:** 在確診數最多的四個縣市，染疫群體的年齡層、性別不同，分析結果反映出各區域的傳染鏈風險。

**關鍵詞：** 新冠肺炎、交互項效應、對數線性模型

Association Among Gender, Age, and Region in Taiwan's First Ten Thousand COVID-19 Cases: A Log-linear-model Analysis

Abstract

**Objectives:** We explore the association between age, gender, and region among Taiwan's 11,290 local Covid-19 cases from January 22, 2020 to June 11, 2021. **Methods:** Using open data from Taiwan's CDC, we organize them into a three-dimensional contingency table. The groups are gender, age 0-29, 30-59, and 60+ years old, and two classifications for region: (1) 7 commonly-defined regions, (2) 12 groups separating Taipei, New Taipei, Keelung, Taoyuan, Hsinchu county, Miaoli county, and Hsinchu city. We adopt the log-linear model for statistical analysis and use the BIC for model selection. **Results:** The model with three pairwise interaction terms has the smallest BIC. In terms of interaction effects, there are more females than males among 30-59 ($p<0.001$), while more males than females among 60+ ($p=0.028$). Miaoli County has more male than female cases ($p<0.001$). Differences between 30-59 and 0-29 (baseline), and between 60+ and 0-29 are significant in Taipei ($p=0.002$ and $p<0.001$); similar age effects for New Taipei is observed; Miaoli County has significant difference between 60+ and 0-29 ($p<0.001$). All Taoyuan's interaction terms are not significant. The main effects of age, the differences between 30-59 and 0-29 (baseline), and between 60+ and 0-29, are both significant ($p=0.002$ and $p=0.046$). **Conclusions:** In the four regions with larger numbers of Covid-19 cases, the age and gender characteristics of the infected population are different, reflecting patterns of infection chains.

# 前言

2020 年世界範圍的 Covid-19 病毒大爆發，已發展成一個國際公共衛生緊急事件，截至 2023 年 3 月 1 日，全球已累計高達超過 6 億確診案例，死亡人數超過 680 萬人[1]。台灣與中國大陸地理位置鄰近，曾被約翰霍普金斯大學預估台灣將會成為疫情的重災區 [2]。幸運的是，台灣採取的防疫措施得宜，至 2020 年年底累積僅有 56 個確診本土案例 [3]，堪稱為防疫的模範生。然而 2021 年 5 月初，台灣本土疫情突發，5 月 19 日防疫警戒由二級提升至三級，各級學校停止到校上課。同年 6 月中之後，台灣本土確診案例逐漸下降，疫情指揮中心 7 月 23 日宣佈 7 月 27 日防疫措施由三級警戒下調降至二級。

本研究是關注台灣 2020 年 1 月 22 日至 2021 年 6 月 11 日的 11290 本土確診案例，多數仍為五月初突發疫情之後的本土案例。我們想了解確診案例中，年齡、性別與所居住地區之間的關係，從而探究是否哪些人群有較高感染風險。一般可以使用單變量分析，去看確診案例中性別分佈、年齡分佈及地區分佈，但是單變量分析方法並不能檢查出它們的交互作用。例如，台北市及新北市的疫情一般猜測是從萬華開始，那麼台北及新北的確診案例是否在 60 歲以上的年齡層較多? 在苗栗縣的確診案例多數是移民工作者，是否 60 歲以上感染者較少? 此為年齡層與縣市地區的交互作用效應 (interaction effects)。我們使用對數線性模型(log-linear models)來探究確診案例中年齡、性別、所居住縣市之間是否有交互作用效應。

在一些針對第一波疫情的國外文獻中，Dowd et al. (2020) [4] 提出了解感染人群中的年齡層有助了解各國 Covid-19 死亡率的差異，Dowd et al. (2020) 引用義大利在 2020 年 3 月 20 日的數據， Covid-19 死亡率在 80 歲以上的是 27.7%，而 44-49 歲則為 0.7%，且 96.9%的死亡發生在 60 歲及以上的人群，因此他們建議防疫政策應隨年齡調整。在義大利雷焦艾米利亞省(Reggio Emilia)2020 年 2 月 27 日-4 月 2 日的人群研究(population study)中[5]， 50 歲以下女性感染率高於同齡男性（2.61% 對 1.84%），但 80 歲以上女性的感染率較男性低（16.49% 對 20.86%）。一項基於美國 24 家醫療機構 2020 年 1 月 20 日-5 月 26 日的 31,461 COVID-19 成人患者的回顧性研究中[6]，患者年齡分布為 18-50 歲 49.5%、50-69 歲 33.8%、70-90 歲 22.3%，女性 54.5%，地區分布為東北部 27%、中西部 21%、南部 30%、西部 21%，該文獻並沒有討論交互作用效應。因多數國外地區疫情嚴重，所以大多國外文獻關注死亡率與解釋變數的關係，一般對染疫人群只作敘述性統計(單變量分析)，沒有探究染疫群體中是否哪些子群有較高感染風險。

## 材料與方法

### 研究動機

在 2021 年 5 月台灣本土疫情突發之時，本文作者希望以自身的統計背景為台灣的防疫工作貢獻一份心力，本文第一作者注意到疾病管制署之開放資料，於是嘗試分析開放資料探尋確診案例中年齡、性別、地區之間有無關聯性

(association)。

## 資料集

本研究使用台灣疾病管制署之「地區年齡性別統計表－嚴重特殊傳染性肺炎－依個案研判日統計」[3]，日期自 2020 年 1 月 22 日至 2021 年 6 月 11 日，並移除境外移入個案，總共確診案例數為 11290，最後總和同年齡、縣市與性別之確診數。本研究進一步合併縣市與年齡的數據，年齡合併的方式以 0-29 歲、30-59 歲、60 歲以上做為組別；縣市合併的方式以北北基、桃竹苗、中彰投、雲嘉南、高屏、宜花東與離島共 7 組做為組別，成為三維度（性別、年齡、縣市地區）的列聯表(表一)。此外，由於台北、新北、桃園、苗栗的確診案例較多，因此也考慮將北北基、桃竹苗拆開的組別方式，只合併中彰投、雲嘉南、高屏、宜花東與離島，共 12 組做為組別。離島雖僅有 9 例，但其具有地理特殊性，不與其他縣市合併。例如美國 CDC 將離島維爾京群島的確診案例數分開計算，不併入離其近的波多黎各。

## 統計分析方法

首先我們使用 R 軟體 4.0.5 版 [7] 及 ggplot2 [8] 畫出各個年齡、縣市、性別之確診人數分佈，然後使用 SAS 統計軟體 9.4 版 [9] 對三維度(X, Y, Z)的列聯表進行對數線性模型(log-linear model) [10] 分析，以查看年齡、性別與縣市關聯性。對數線性模型屬於廣義線性模型 (generalized linear model) [11]，它先將列聯表中每個格子計數(cell counts)視為獨立卜瓦松分佈(Poisson distribution)，其相應均值 (means)等於格子計數期望值(expected cell counts)$\mu_{ijk}$，假設總確診數 11290 為給定的條件下，列聯表中每個格子計數為機率$\pi_{ijk}$的多項式分佈(multinomial distribution) [10]，在對數線性模型中以 log($\pi_{ijk}$)(等同 log($\mu_{ijk}$))為反應變數，性別、年齡與縣市為解釋變數。儘管原始的多項式分佈假設案例彼此之間獨立，而一些染疫個案間可能不獨立(如家戶傳染)，一般在大樣本之下，獨立性假設是合理的。且在對數線性模型下，探究的是案例落在格子機率$\pi_{ijk}$和性別、年齡與縣市有無關聯(主效應(main effects))以及性別、年齡與縣市之間關聯性(交互作用效應)，獨立假設並不意味著機率$\pi_{ijk}$之間沒有關聯性。

在對數線性模型中以男性、0-29 歲與宜花東作為基準(baseline)。模型建構方式如以下步驟，第一，放入性別、年齡與縣市之主效應(main effects)，此模型隱喻這三個變量是相互獨立的(mutual independent)；第二，放入主效應變數後，再放入一個成對交互項(interaction term)，此模型隱喻在成對交互項中的兩個變量與第三個變量是共同獨立的(jointly independent)；第三，放入主效應變數後，再放入兩個成對交互項 X*Z 及 Y*Z，此模型隱喻 X 及 Y 在給定 Z 的條件下是條件獨立的(conditional independent)；第四，放入主效應變數後，再放入三個成對交互項(pairwise interaction terms)，此模型隱喻這三個變量是同質關聯的(homogeneous association)；第五，放入主效應變數及成對交互項後，再放入三變數之交互項，此模型具有列聯表中總格子數一樣多的參數，是飽和的且自由度為 0 [10]。最後計算以上各個模型之 BIC [12]，並使用 BIC 做為模型選擇指標，

統計檢定以 p 值小於 0.05 來檢視性別、年齡與縣市之間是否有關聯性。

結果

### 單變量分析(Univariate Analysis)

性別方面，男性確診人數(n= 5586, 49.5%)跟女性確診人數(n= 5704, 50.5%)並沒有太大的差異；而年齡方面，不同年齡層之確診人數有差異。其中，最多人數是落在 30-59 歲(n= 5492, 48.6%)，其次是 60 歲以上(n= 4046, 35.8%)，最後是 29 歲以下(n= 1752, 15.5%)；縣市方面，不同區域的確診人數有明顯的差異，其中最多人數為北北基(n=9453, 83.7%)，其次是桃竹苗(n=1041, 9.2%)，再來是中彰投(n=450, 4.0%)，如圖一 a 所示。此外，在北北基中，確診數最多的是新北市(n=5423)，其次是台北市(n=3804)，然後是基隆市(n=226)；而在桃竹苗中，確診數最多的是桃園市(n=549)，其次是苗栗縣(n=409)、新竹縣(n=50)、新竹市(n=33)，如圖二 a 所示。

### 對數線性模型分析

基於表一的三維度列聯表，總共 9 個對數線性模型及其對應之 BIC 如表二所示，其中 BIC 最小的是包含三個成對交互項的模型，因此該模型作為最終模型。以下模型一的結果是基於表一的含有三個成對交互項的對數線性模型，模型二則是將北北基、桃竹苗拆開的含有成對交互項的對數線性模型，其 BIC 在 9 個模型中也為最小(表二)。含有三個成對交互項的對數線性模型隱喻這三個變

量是同質關聯的[10]，意即任何兩個變量之間的條件優勢(conditional odds ratios)在給定第三個變量的每個類別中都是相同的(conditional odds ratios between any two variables are identical at each category of the third variable)。

**模型一**

模型一之參數估計值如表三所示，因為一些交互項是統計上顯著的，在這種情況下，同時考量交互作用項效應和主效應。解釋如下。

1. 性別*縣市方面，只有桃竹苗*性別是顯著的(p=0.003)，代表在桃竹苗的估計值男性比女性多，如圖一 b 所示。

2. 年齡*性別方面，(30-59 歲*女性)及(60 歲以上*女性)皆為顯著(p<0.001 及 p=0.028)，代表在 30-59 歲的估計值女性比男性多。60 歲以上則是男性比女性多，如圖一 c 所示。年齡主效應 0-59 歲、60 歲以上與 0-29 歲(基準)的差異皆為顯著(p=0.002 及 p=0.046)，且 30-59 歲及 60 歲的估計值皆大於 0-29 歲，從估計數值來說，最多的是 30-59 歲。

3. 年齡*縣市方面，如圖一 d 所示，在 30-59 歲的年齡層，只有北北基交互項是顯著的(p=0.018)，代表在北北基的估計值，29 歲以下的比 30-59 歲的少。而 60 歲以上的年齡層，北北基、桃竹苗的交互項是顯著的(p=0.003 及 p<0.001)，代表在北北基的估計值，60 歲以上的比 29 歲以下的多，而桃竹苗的估計值，60 歲以上的比 29 歲以下的少。比較(北北基*年齡層)交互項的估計值，我們發現 60 歲以上的估計值大於 30-59 歲，然而，如果使用 30-59 歲作為基準，發現北北基 60 歲以上的估計值與 30-59 歲的估計值之間是不顯著(p=0.354)。縣市主效應方面，北北基、桃竹苗、中彰投、雲嘉南、離島與宜花東(基準)差異是顯著的，從估計數值來說，代表大於宜花東的有北北基、桃竹苗及中彰投，其中最大的是北北基，其次是桃竹苗，最後是中彰投。反之，雲嘉南及離島的估計值小於宜花東。

**模型二**

模型二是將北北基、桃竹苗拆開的含有三個成對交互項的對數線性模型，由於模型二在性別與年齡之交互項效應與模型一類似，所以模型二只討論北北基、桃竹苗縣市之性別*縣市及年齡*縣市效應，模型二之參數估計值如表四所示，解釋如下。

1. 性別*縣市方面，苗栗縣*性別有顯著差異(p<0.001)，代表在苗栗縣的估計值，男性比女性多，如圖二 b 所示。

2. 年齡*縣市方面，在 30-59 歲的年齡層，台北市有顯著差異(p=0.002)，新北市(p=0.062)接近顯著，代表在新北市及台北市的估計值，29 歲以下的比 30-59 歲的少。而在 60 歲以上的年齡層，苗栗縣(p<0.001)、新北市(p=0.018)、台北市(p<0.001)有顯著差異，代表在新北市、台北市的估計值，60 歲以上的比 29 歲以下的多；而苗栗縣的估計值，60 歲以上的比 29 歲以下的少，如圖二 c 所示。比較台北市*年齡層交互項的估計數值，我們發現 60 歲以上的估計值大於 30-59 歲的估計值，且兩者皆顯著多於 0-29 歲(基準)。然而，如果使用 30-59 歲作為基

準，發現台北市 60 歲以上的估計值與 30-59 歲之間是不顯著(p=0.220)。同理，比較新北市*年齡層交互項的估計數值，發現 60 歲以上的估計值大於 30-59 歲的估計值，且兩者皆顯著多於 0-29 歲(基準)。然而，如果使用 30-59 歲作為基準，發現新北市 60 歲以上的估計值與 30-59 歲之間不顯著(p=0.443)。而苗栗縣 60 歲以上的估計值顯著小於 30-59 歲的估計值(p<0.001) 。在縣市主效應方面，新竹市、新竹縣、苗栗縣、新北市、台北市、桃園市與宜花東(基準)差異是顯著的，估計值最多的是新北市，其次為台北市、苗栗縣及桃園市，最後是中彰投。反之小於宜花東的估計值有新竹縣、新竹市、雲嘉南及離島。

## 討論

對於三維列聯表，對數線性模型提供了一種系統的方法來找尋可能的關聯。本研究的結果顯示，台灣這波疫情主要影響 30 歲以上的群體，達確診案例中近 85%，在 30-59 歲的估計值，女性比男性多。60 歲以上則是男性比女性多。在確診數最多的新北市及台北市，最多案例的年齡層是 60 歲以上，但 60 歲以上的估計值並不顯著大於 30-59 歲。在確診數第三多的桃園市，交互項並不顯著，表示其確診案例中性別或年齡層之間並無顯著差異。確診數第四多的苗栗縣，60 歲以上的估計值顯著小於 30-59 歲及 29 歲以下，且男性比女性多。可以說在確診數最多的四個縣市，染疫群體的年齡層、性別各有不同，分析結果反映出各區域的傳染鏈風險，新北及台北的萬華茶藝館群聚感染，桃園的飛行機組人員及機場防疫旅館感染，苗栗的電子廠移工群聚感染，其中尤以新北及台北的

確診案例最多。前副總統陳建仁曾指出[10],「此前日本風俗業就爆發過疫情,可惜我國未記取日本的前車之鑑」,突顯防疫工作中道德教育的重要性。

因多數國外地區疫情嚴重,所以大多國外文獻關注死亡率與解釋變數的關係,如在前言中提到義大利 2020 年 3 月 20 日的研究[4],文獻一般對染疫人群只作敘述性統計(單變量分析),沒有探究交互作用效應。本研究是在 2021 年 5 月台灣本土疫情突發之後,我們想了解確診案例中,年齡、性別與所居住地區之間有無關聯性,可以說是在當時的時事背景下所衍生的研究動機,與國外疫情嚴重地區的研究動機不同,據我們目前所知,沒有可直接參照和比較的國外研究文獻。

本研究可能的局限性如下: 所使用的數據[3]收集原始、alpha 病毒株在台灣擴散的情形,與現今國際上流行之 Delta 及 Omicron 病毒株傳播速度可能不同,預期指揮中心將超前布署因應變種病毒株之政策,人民也須持續做好防疫相關措施以預防新種病毒株擴散甚至變異。此外,該數據[3]僅能收集到有參與篩檢之確診案例,因此無法估計隱性染疫者造成的影響;加上對數指數模型僅使用各個縣市、年齡層及性別之確診案例,未使用個案研判日期,所以不能判斷病毒在特定區域的傳播週期。

## 參考文獻

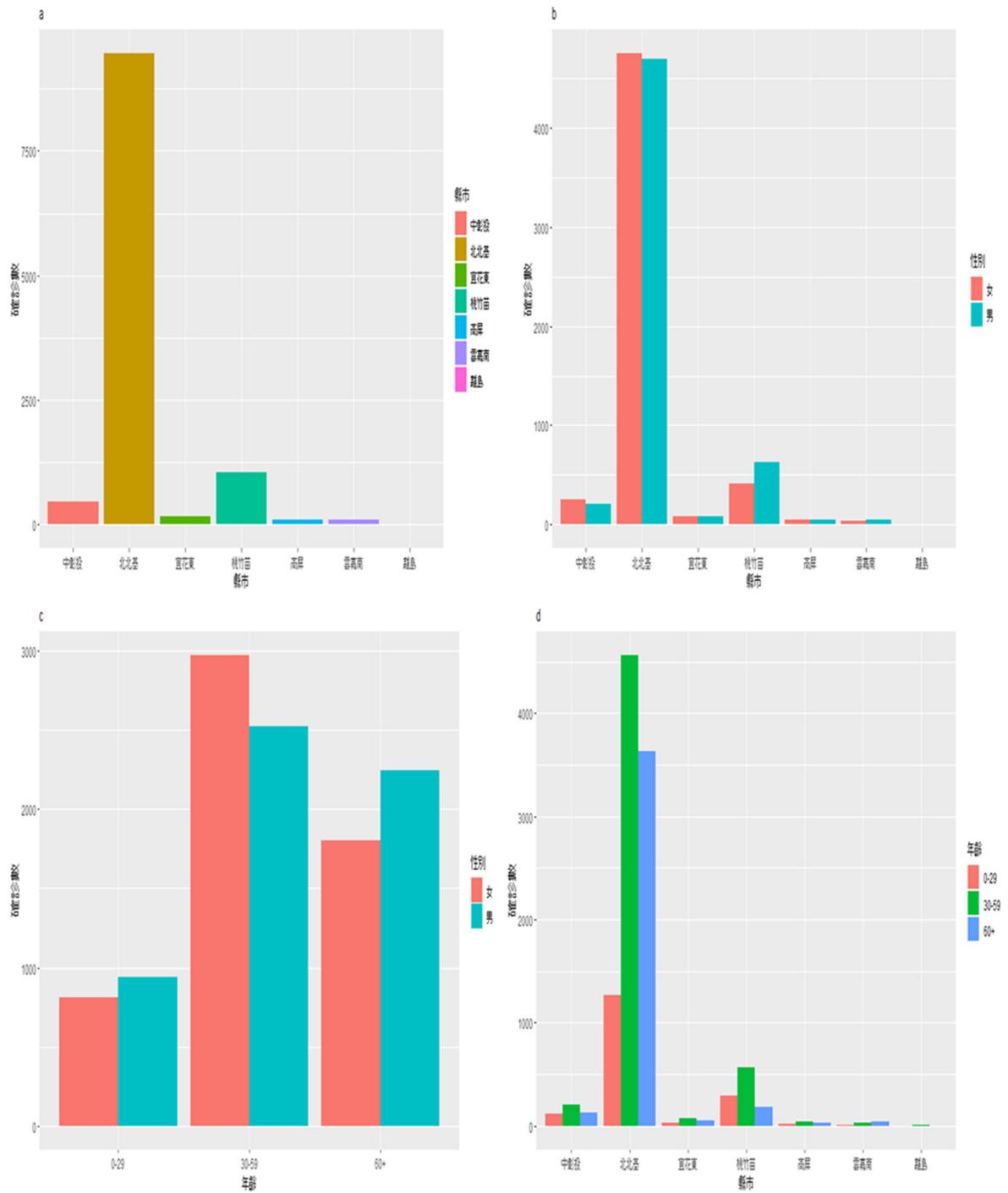

圖一 (a)不同縣市之確診數; (b)不同縣市及性別之確診數; (c)不同年齡及性別之確診數 (d)不同縣市及年齡層之確診數

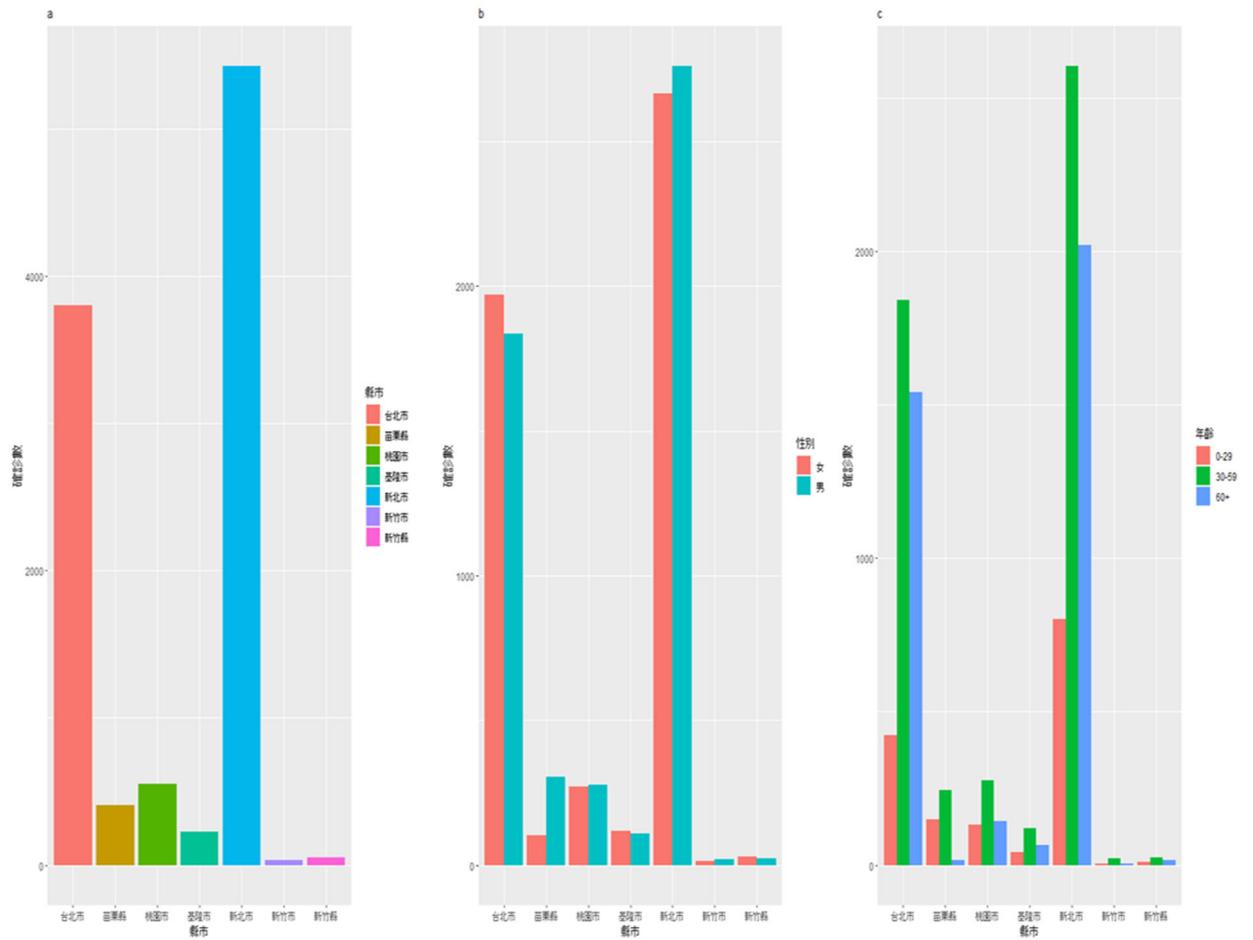

圖二 (a) 北北基、桃竹苗不同縣市之確診數 (b) 北北基、桃竹苗不同縣市及性別之確診數 (c) 北北基、桃竹苗不同縣市及年齡層之確診數

表一 性別*年齡*縣市之列聯表

| 確診案例數 | | 縣市 | | | | | | | |
|---|---|---|---|---|---|---|---|---|---|
| 性別 | 年齡 | 高屏 | 離島 | 桃竹苗 | 中彰投 | 北北基 | 雲嘉南 | 宜花東 | 總計 |
| 女性 | 0-29 歲 | 11 | 1 | 112 | 58 | 615 | 5 | 12 | 814 |
| | 30-59 歲 | 24 | 3 | 226 | 126 | 2530 | 15 | 48 | 2972 |
| | 60 歲以上 | 12 | 1 | 79 | 63 | 1607 | 16 | 22 | 1800 |
| 男性 | 0-29 歲 | 10 | 2 | 184 | 62 | 650 | 8 | 22 | 938 |
| | 30-59 歲 | 22 | 1 | 339 | 80 | 2032 | 19 | 27 | 2520 |
| | 60 歲以上 | 15 | 1 | 101 | 61 | 2019 | 21 | 28 | 2246 |
| | 總計 | 94 | 9 | 1041 | 450 | 9453 | 84 | 159 | 11290 |

表二 建構模型一和模型二所考慮的 9 個對數線性模型及其 BIC，

| | 模型一 | | | 模型二(北北基、桃竹苗拆開) | | |
|---|---|---|---|---|---|---|
| 模型 | BIC | Deviance | 自由度 | BIC | Deviance | 自由度 |
| 年齡+性別+縣市 | 735.9501 | 486.2946 | 32 | 1186.6578 | 743.0137 | 57 |
| 年齡+性別+縣市+年齡*縣市 | 480.0381 | 168.1539 | 20 | 811.2533 | 273.5226 | 35 |
| 年齡+性別+縣市+年齡*性別 | 666.7458 | 392.2382 | 30 | 1101.1549 | 648.9573 | 55 |
| 年齡+性別+縣市+性別*縣市 | 728.9761 | 439.5178 | 26 | 1116.4327 | 625.7453 | 46 |
| 年齡+性別+縣市+年齡*性別+性別*縣市 | 642.3951 | 345.4615 | 24 | 1030.9297 | 531.6889 | 44 |
| 年齡+性別+縣市+年齡*性別+年齡*縣市 | 393.4571 | 74.0975 | 18 | 725.7503 | 179.4462 | 33 |
| 年齡+性別+縣市+年齡*縣市+性別*縣市 | 455.6874 | 121.3771 | 14 | 741.0283 | 156.2542 | 24 |
| 年齡+性別+縣市+年齡*性別+年齡*縣市+性別*縣市 | 361.1369 | 19.3513 | 12 | 639.4719 | 46.1445 | 22 |
| 年齡+性別+縣市+年齡*性別+年齡*縣市+性別*縣市+年齡*性別*縣市 | 386.6376 | 0 | 0 | 687.4141 | 0 | 0 |

表三 基於表一的含有三個成對交互項的對數線性模型(模型一)的結果

| 參數 | | | 估計值 | 標準差 | p.value |
|---|---|---|---|---|---|
| Intercept | | | 2.8483 | 0.1897 | <.0001 |
| 性別 | 女 | | -0.03 | 0.1647 | 0.8539 |
| 年齡 | 30-59 | | 0.6411 | 0.209 | 0.0022 |
| 年齡 | 60+ | | 0.4467 | 0.2241 | 0.0462 |
| 縣市 | 高屏 | | -0.447 | 0.3047 | 0.1424 |
| 縣市 | 離島 | | -2.508 | 0.705 | 0.0004 |
| 縣市 | 桃竹苗 | | 2.3823 | 0.1988 | <.0001 |
| 縣市 | 中彰投 | | 1.1933 | 0.2154 | <.0001 |
| 縣市 | 北北基 | | 3.6392 | 0.191 | <.0001 |
| 縣市 | 雲嘉南 | | -0.817 | 0.3477 | 0.0188 |
| 性別*縣市 | 女 | 高屏 | -0.072 | 0.2614 | 0.7829 |
| 性別*縣市 | 女 | 離島 | 0.1573 | 0.6919 | 0.8202 |
| 性別*縣市 | 女 | 桃竹苗 | -0.508 | 0.1717 | 0.0031 |
| 性別*縣市 | 女 | 中彰投 | 0.1333 | 0.1856 | 0.4727 |
| 性別*縣市 | 女 | 北北基 | -0.047 | 0.1608 | 0.771 |
| 性別*縣市 | 女 | 雲嘉南 | -0.318 | 0.2729 | 0.2433 |
| 年齡*性別 | 30-59 | 女 | 0.2842 | 0.0555 | <.0001 |
| 年齡*性別 | 60+ | 女 | -0.128 | 0.0584 | 0.0281 |
| 年齡*縣市 | 30-59 | 高屏 | -0.002 | 0.3353 | 0.9955 |
| 年齡*縣市 | 30-59 | 離島 | -0.515 | 0.7927 | 0.5163 |
| 年齡*縣市 | 30-59 | 桃竹苗 | -0.109 | 0.2193 | 0.6191 |
| 年齡*縣市 | 30-59 | 中彰投 | -0.26 | 0.2369 | 0.2721 |
| 年齡*縣市 | 30-59 | 北北基 | 0.4949 | 0.2095 | 0.0182 |
| 年齡*縣市 | 30-59 | 雲嘉南 | 0.1928 | 0.3866 | 0.618 |
| 年齡*縣市 | 60+ | 高屏 | -0.137 | 0.3662 | 0.7091 |
| 年齡*縣市 | 60+ | 離島 | -0.786 | 0.9398 | 0.4029 |
| 年齡*縣市 | 60+ | 桃竹苗 | -0.899 | 0.2417 | 0.0002 |
| 年齡*縣市 | 60+ | 中彰投 | -0.349 | 0.2566 | 0.1743 |
| 年齡*縣市 | 60+ | 北北基 | 0.6659 | 0.2247 | 0.003 |
| 年齡*縣市 | 60+ | 雲嘉南 | 0.6503 | 0.3917 | 0.0969 |

表四 將北北基、桃竹苗拆開的含有三個成對交互項的對數線性模型(模型二)的結果

| 參數 | | | 估計值 | 標準差 | p.value | 參數 | | | 估計值 | 標準差 | p.value |
|---|---|---|---|---|---|---|---|---|---|---|---|
| Intercept | | | 2.8379 | 0.1901 | <.0001 | 年齡*性別 | 30-59 | 女 | 0.2665 | 0.0559 | <.0001 |
| 性別 | 女 | | -0.0095 | 0.1648 | 0.9542 | 年齡*性別 | 60+ | 女 | -0.1679 | 0.0588 | 0.0043 |
| 年齡 | 30-59 | | 0.6497 | 0.209 | 0.0019 | 年齡*縣市 | 30-59 | 新竹市 | 0.4987 | 0.5077 | 0.326 |
| 年齡 | 60+ | | 0.4657 | 0.2242 | 0.0378 | 年齡*縣市 | 30-59 | 新竹縣 | 0.2191 | 0.4408 | 0.6191 |
| 縣市 | 新竹市 | | -1.4932 | 0.468 | 0.0014 | 年齡*縣市 | 30-59 | 高屏 | -0.0022 | 0.3353 | 0.9948 |
| 縣市 | 新竹縣 | | -1.4191 | 0.4133 | 0.0006 | 年齡*縣市 | 30-59 | 基隆市 | 0.2751 | 0.2752 | 0.3174 |
| 縣市 | 高屏 | | -0.4463 | 0.3053 | 0.1438 | 年齡*縣市 | 30-59 | 苗栗縣 | -0.2297 | 0.2321 | 0.3223 |
| 縣市 | 基隆市 | | 0.1924 | 0.2539 | 0.4486 | 年齡*縣市 | 30-59 | 新北市 | 0.3931 | 0.2109 | 0.0624 |
| 縣市 | 苗栗縣 | | 1.9171 | 0.2078 | <.0001 | 年齡*縣市 | 30-59 | 離島 | -0.5135 | 0.7925 | 0.517 |
| 縣市 | 新北市 | | 3.2035 | 0.1927 | <.0001 | 年齡*縣市 | 30-59 | 中彰投 | -0.2594 | 0.2368 | 0.2734 |
| 縣市 | 離島 | | -2.5069 | 0.7066 | 0.0004 | 年齡*縣市 | 30-59 | 台北市 | 0.6782 | 0.2139 | 0.0015 |
| 縣市 | 中彰投 | | 1.1935 | 0.2158 | <.0001 | 年齡*縣市 | 30-59 | 桃園市 | -0.043 | 0.2327 | 0.8534 |
| 縣市 | 台北市 | | 2.5106 | 0.1958 | <.0001 | 年齡*縣市 | 30-59 | 雲嘉南 | 0.1912 | 0.3866 | 0.6209 |
| 縣市 | 桃園市 | | 1.3968 | 0.2121 | <.0001 | 年齡*縣市 | 60+ | 新竹市 | -0.4084 | 0.6189 | 0.5094 |
| 縣市 | 雲嘉南 | | -0.8171 | 0.3483 | 0.019 | 年齡*縣市 | 60+ | 新竹縣 | 0.197 | 0.4725 | 0.6768 |
| 性別*縣市 | 女 | 新竹市 | -0.5639 | 0.3915 | 0.1498 | 年齡*縣市 | 60+ | 高屏 | -0.1374 | 0.3663 | 0.7076 |
| 性別*縣市 | 女 | 新竹縣 | 0.1733 | 0.3277 | 0.5969 | 年齡*縣市 | 60+ | 基隆市 | 0.09 | 0.2984 | 0.763 |
| 性別*縣市 | 女 | 高屏 | -0.0728 | 0.2614 | 0.7806 | 年齡*縣市 | 60+ | 苗栗縣 | -2.7328 | 0.3508 | <.0001 |
| 性別*縣市 | 女 | 基隆市 | -0.0105 | 0.2081 | 0.9597 | 年齡*縣市 | 60+ | 新北市 | 0.5355 | 0.2263 | 0.0179 |
| 性別*縣市 | 女 | 苗栗縣 | -1.2242 | 0.1962 | <.0001 | 年齡*縣市 | 60+ | 離島 | -0.7847 | 0.94 | 0.4038 |
| 性別*縣市 | 女 | 新北市 | -0.0905 | 0.1618 | 0.5757 | 年齡*縣市 | 60+ | 中彰投 | -0.3474 | 0.2567 | 0.1759 |
| 性別*縣市 | 女 | 離島 | 0.1532 | 0.6921 | 0.8248 | 年齡*縣市 | 60+ | 台北市 | 0.9074 | 0.2291 | <.0001 |
| 性別*縣市 | 女 | 中彰投 | 0.1316 | 0.1857 | 0.4786 | 年齡*縣市 | 60+ | 桃園市 | -0.3021 | 0.2532 | 0.2327 |
| 性別*縣市 | 女 | 台北市 | 0.0207 | 0.1628 | 0.8987 | 年齡*縣市 | 60+ | 雲嘉南 | 0.6473 | 0.3918 | 0.0985 |
| 性別*縣市 | 女 | 桃園市 | -0.0988 | 0.181 | 0.5853 | | | | | | |
| 性別*縣市 | 女 | 雲嘉南 | -0.3148 | 0.273 | 0.2488 | | | | | | |

20